\documentclass[sigconf]{acmart}
\AtBeginDocument{%
  }

\setcopyright{acmlicensed}
\copyrightyear{2025}
\acmYear{2026}
\acmDOI{XXXXXXX.XXXXXXX}
\acmConference[TSMO Workshop in conjunction with KDD]{TSMO Workshop in conjunction with KDD}{August 9, 2026}{Jeju, Korea}
\acmISBN{978-1-4503-XXXX-X/2018/06}

\usepackage{tikz}
\usetikzlibrary{shapes.geometric, arrows, positioning, calc, backgrounds, fit, decorations.pathreplacing}
\tikzset{
  decision/.style={diamond, draw, text width=4.5em, text badly centered, inner sep=0pt},
  block/.style={rectangle, draw, text width=10em, text centered, rounded corners, minimum height=3em},
  line/.style={draw, -latex'},
  cloud/.style={draw, ellipse, minimum height=2em},
  note/.style={rectangle, dashed, draw, inner sep=5pt}
}
\usepackage{tcolorbox}
\usepackage{booktabs}
\usepackage{epigraph}
\begin{document}

\title{Recommending Search Filters To Improve Conversions At Airbnb}

\author{Hao Li}
\email{hao.li@airbnb.com}
\affiliation{%
  \institution{Airbnb, Inc.}
  \country{USA}
}

\author{Kedar Bellare}
\email{kedar.bellare@airbnb.com}
\affiliation{%
  \institution{Airbnb, Inc.}
  \country{USA}
}

\author{Siyu Yang}
\email{siyu.yang@airbnb.com}
\affiliation{%
  \institution{Airbnb, Inc.}
  \country{USA}
}

\author{Sherry Chen}
\email{sherry.chen@airbnb.com}
\affiliation{%
  \institution{Airbnb, Inc.}
  \country{USA}
}

\author{Liwei He}
\email{liwei.he@airbnb.com}
\affiliation{%
  \institution{Airbnb, Inc.}
  \country{USA}
}

\author{Stephanie Moyerman}
\email{stephanie.moyerman@airbnb.com}
\affiliation{%
  \institution{Airbnb, Inc.}
  \country{USA}
}

\author{Sanjeev Katariya}
\email{sanjeev.katariya@airbnb.com}
\affiliation{%
  \institution{Airbnb, Inc.}
  \country{USA}
}

\renewcommand{\shortauthors}{Hao Li et al.}

\begin{abstract}
Airbnb, a two-sided online marketplace connecting guests and hosts, offers a diverse and unique inventory of accommodations, experiences, and services. Search filters play an important role in helping guests navigate this variety by refining search results to align with their needs. Yet, while search filters are designed to facilitate conversions in online marketplaces, their direct impact on driving conversions remains underexplored in the existing literature. 

This paper bridges this gap by presenting a novel application of machine learning techniques to recommend search filters aimed at improving booking conversions. We introduce a modeling framework that directly targets lower-funnel conversions (bookings) by recommending intermediate tools, i.e. search filters. Leveraging the framework, we designed and built the filter recommendation system at Airbnb from the ground up, addressing challenges like cold start and stringent serving requirements. 

The filter recommendation system we developed has been successfully deployed at Airbnb, powering multiple user interfaces and driving incremental booking conversion lifts, as validated through online A/B testing. An ablation study further validates the effectiveness of our approach and key design choices. By focusing on conversion-oriented filter recommendations, our work ensures that search filters serve their ultimate purpose at Airbnb — helping guests find and book their ideal accommodations.
\end{abstract}




\maketitle

\section{Introduction}
In online marketplaces, search filters are essential tools that allow users to narrow down search results by specifying their criteria. Airbnb offers a diverse and unique inventory of accommodations, each with its own distinct characteristics. While this variety enhances the appeal of the platform, it can make finding the perfect stay challenging for guests. As the most used search tool, search filters streamline the search process, enabling guests to refine their search criteria and quickly identify listings aligned with their preferences.

\begin{figure}[t]
    \centering
    \includegraphics[width=0.5\linewidth]{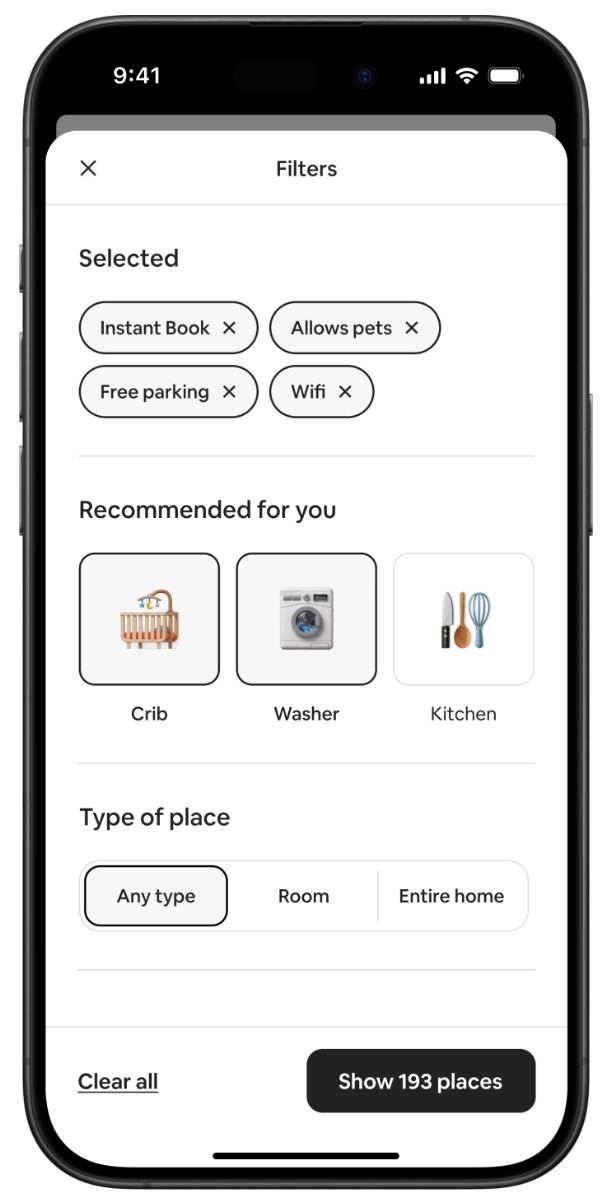}
    \caption{Recommending filters in the “Recommended for you” section in the filter panel}
    \label{fig:recommended-for-you}
\end{figure}

\begin{figure}[t]
    \centering
    \includegraphics[width=0.5\linewidth]{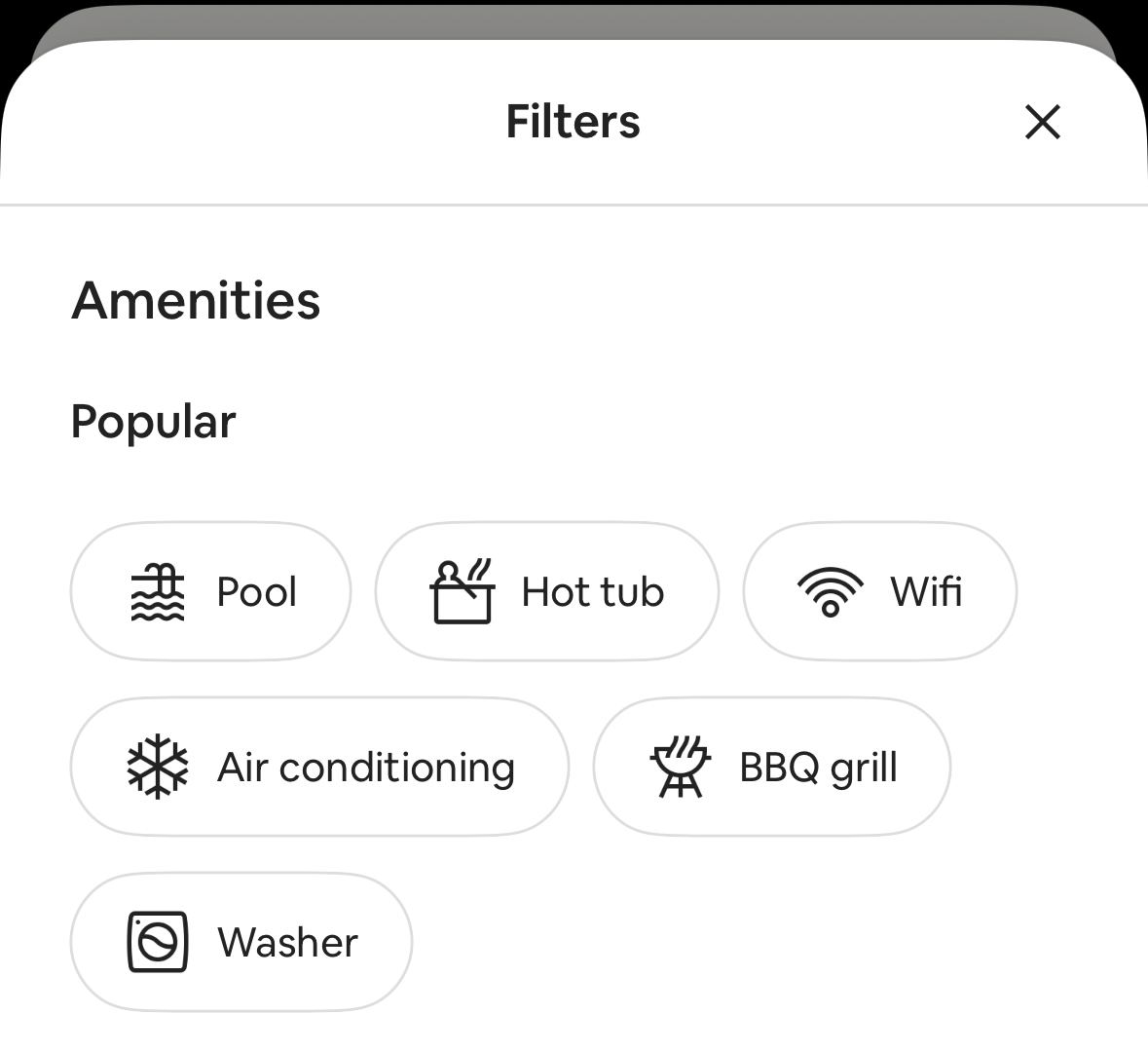}
    \caption{Recommending amenity filters in the amenities section in the filter panel}
    \label{fig:recommended-amenities}
\end{figure}

\begin{figure}[t]
    \centering
    \includegraphics[width=1.0\linewidth]{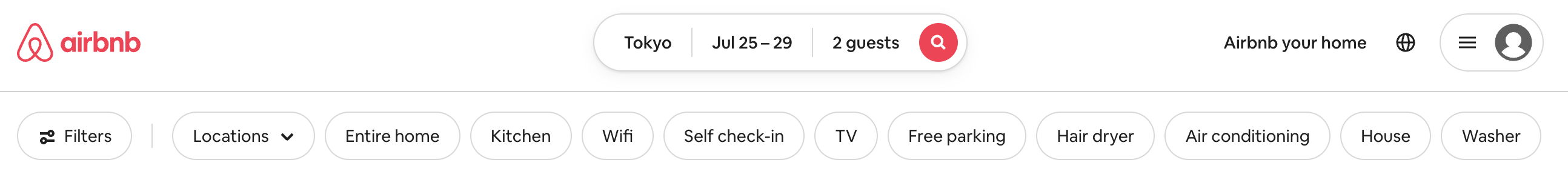}
    \caption{Recommending filters in the filter bar}
    \label{fig:filter-bar}
\end{figure}

Recommending search filters dynamically is important for filter discoverability. On the demand side, guest needs vary, and the filter panel is extensive and can be overwhelming. On the supply side, Airbnb has various types of listings that cater to different needs, which guests might not be familiar with. For example, Airbnb provides a “guest favorite" filter for travelers prioritizing quality, “dedicated workspace” for digital nomads and “allows pets” for pet owners. Recommending search filters addresses these challenges by boosting filter discoverability on both sides: (1) helping guests find the filters they need and (2) increasing awareness of useful filters. Figure~\ref{fig:recommended-for-you}, Figure~\ref{fig:recommended-amenities} and Figure~\ref{fig:filter-bar} illustrate the UI surfaces where search filter recommendations are displayed.

While enhancing filter discoverability is valuable, it addresses only part of the equation. Search filters are tools, not the end goal. For e-commerce platforms and two-sided online marketplaces, the primary focus is on driving conversions. This means going beyond guest engagement with search filters to ensuring successful progress toward conversions. Filters are merely middle steps in the journey guests undertake to achieve the goal of using Airbnb — booking the best accommodation for their upcoming trips. Why settle for better discoverability when we should ensure the tools serve their purpose? 

This brings us to the key question: \textbf{can recommending search filters improve conversions?}

To answer this question, we introduce a novel application of machine learning techniques to recommend search filters to improve conversions. Our contributions are threefold:
\begin{itemize}
\item {\textbf{Framework}}: We propose a modeling framework that optimizes lower-funnel conversions by recommending intermediate tools, i.e. search filters, rather than end items like listings. This includes a unique problem formulation, problem decomposition and training example construction. The framework is designed to be domain-agnostic, applicable to online marketplaces where filters or facets are used to guide users toward a final conversion goal.
\item {\textbf{System}}: We detail the system design and practical considerations required to productionize the framework. This includes cold start strategies to bootstrap the system during the zero-to-one stage and a scalable, extensible architecture.
\item {\textbf{Empirical validation}}: We provide evidence through online A/B testing and ablation study, demonstrating the effectiveness of the proposed framework and system in directly driving conversions via search filter recommendations.
\end{itemize}

While this study focuses on Airbnb, the challenges of matching diverse inventory with user-specific needs and achieving the business goal of increasing conversions are common to online marketplaces, making our approach broadly applicable to other platforms.

The rest of the paper is organized as follows: Section 2 reviews related work on faceted search and recommender systems for conversion optimization. Section 3 motivates the study through exploratory data analysis, showcasing why the problem is worth addressing and providing intuitive examples to justify the investment. Section 4 introduces our modeling framework, covering the problem formulation, example construction, and model architecture. Section 5 discusses the system design and practical considerations for productionization. Section 6 presents the results of offline and online experiments alongside an ablation study. Section 7 concludes the paper and highlights opportunities for future work.

\section{Related Work}
Numerous prior studies have been conducted to address the same problem -- ranking filters in faceted search systems, yet our approach differs from them in the end goal. They primarily focus on improving search navigation in general, while we focus on improving conversions which is a common business goal in e-commerce. There are other previous works on recommender systems and conversion optimization, yet our approach differs from them in what to recommend. They focus on recommending the end item while we focus on the search filters, which are intermediate search tools. 

\textbf{Ranking Filters in Faceted Search Systems}.
Search filters, a key component of faceted search systems, help users narrow down search results based on specific criteria. Prior works on facet/filter ranking primarily aim at improving navigational ease or user engagement. Survey papers, such as \cite{wei2013survey} and \cite{ali2023comprehensive}, provide overviews of faceted search and facet ranking methodologies. Specific approaches include probabilistic generative framework proposed by \cite{koren2008personalized}\, as well as efforts to reduce navigational cost, typically measured by the number of clicks to find the desired product, by \cite{vandic2013facet} and \cite{vandic2017dynamic}. Other works \cite{bernardi2020recommending}\cite{pradhan2023dynamic} incorporate user feedback (e.g., filter engagement) but do not explicitly target conversion-based outcomes. Additionally, the methods proposed in \cite{koren2008personalized}\cite{vandic2013facet}\cite{vandic2017dynamic} are evaluated using small datasets, simulations or user studies; while those in \cite{bernardi2020recommending}\cite{pradhan2023dynamic} are evaluated by running online A/B tests but targeting upper-funnel metrics like filter engagement.

In contrast, our work goes beyond navigation or filter engagement. We directly target lower-funnel conversions through filter recommendations, validated through online A/B testing.

\textbf{Recommender Systems and Search Ranking for Conversion Optimization}.
Recommender systems and search ranking have been widely studied and deployed to optimize diverse business metrics. For example, social media platforms like YouTube \cite{covington2016deep} and Pinterest \cite{pal2020pinnersage}\cite{pancha2022pinnerformer} optimize for upper-funnel engagement metrics such as watch time and interactions. Ad and e-commerce platforms, such as Meta \cite{he2014practical}, Microsoft \cite{shan2016deep}\cite{huang2013learning}, and Alibaba \cite{zhou2018deep}, often emphasize click-through rates. However, Airbnb focuses more on the lower-funnel conversion, specifically uncanceled bookings. Unlike platforms where increased engagement alone is valuable, improving engagement at Airbnb without driving conversions would degrade the search experience. In the context of optimizing conversions, e-commerce and marketplace platforms have extensively studied search ranking and recommendation. Examples include Google Play for app install \cite{cheng2016wide}, Alibaba for online shopping conversions \cite{ma2018entire}, and Airbnb for accommodation bookings \cite{grbovic2018real}\cite{haldar2019applying}\cite{haldar2020improving}\cite{haldar2023learning}\cite{tan2023optimizing}\cite{tang2024multi}. However, prior work predominantly focuses on recommending end items (e.g., apps, products, or listings) where conversions occur.

In contrast, our work uniquely targets intermediate tools, search filters. By recommending search filters to improve conversions, we bridge the gap between traditional recommender systems and search ranking to address a novel, business-critical use case.

\section{Motivating Example}
\begin{quote}
\textit{Can recommending search filters improve conversions?}
\end{quote}

We now address the key question posed in the introduction. To answer this question, we start by examining the search journey to understand how filter usage impacts booking conversions at Airbnb –  it’s a two-step process: 
\begin{enumerate}
    \item guests decide whether to apply certain filters.
    \item guests choose whether to book a listing from the filtered results.
\end{enumerate}

The key hypothesis of this paper lies in the second step: if a filter is likely to contain a listing that the guest will book, then recommending that filter for selection can improve booking conversion by making the desired  listing more discoverable. 

We define Filter Conversion Rate (FCR) to measure the second part:
\begin{equation}
\text{FCR} = 
\frac{N_{\text{converted, filter applied}}}
{N_{\text{filter applied}}}
\label{eq:filter-conversion-rate}
\end{equation}

We define the terms in the equation as follows:
\begin{itemize}
    \item \( N_{\text{filter applied}} \): The number of searches where the filter is applied.
    \item \( N_{\text{converted, filter applied}} \): The number of searches where the filter is applied and leads to conversion.
\end{itemize}

\begin{figure}[t]
    \centering
    \includegraphics[width=0.8\linewidth]{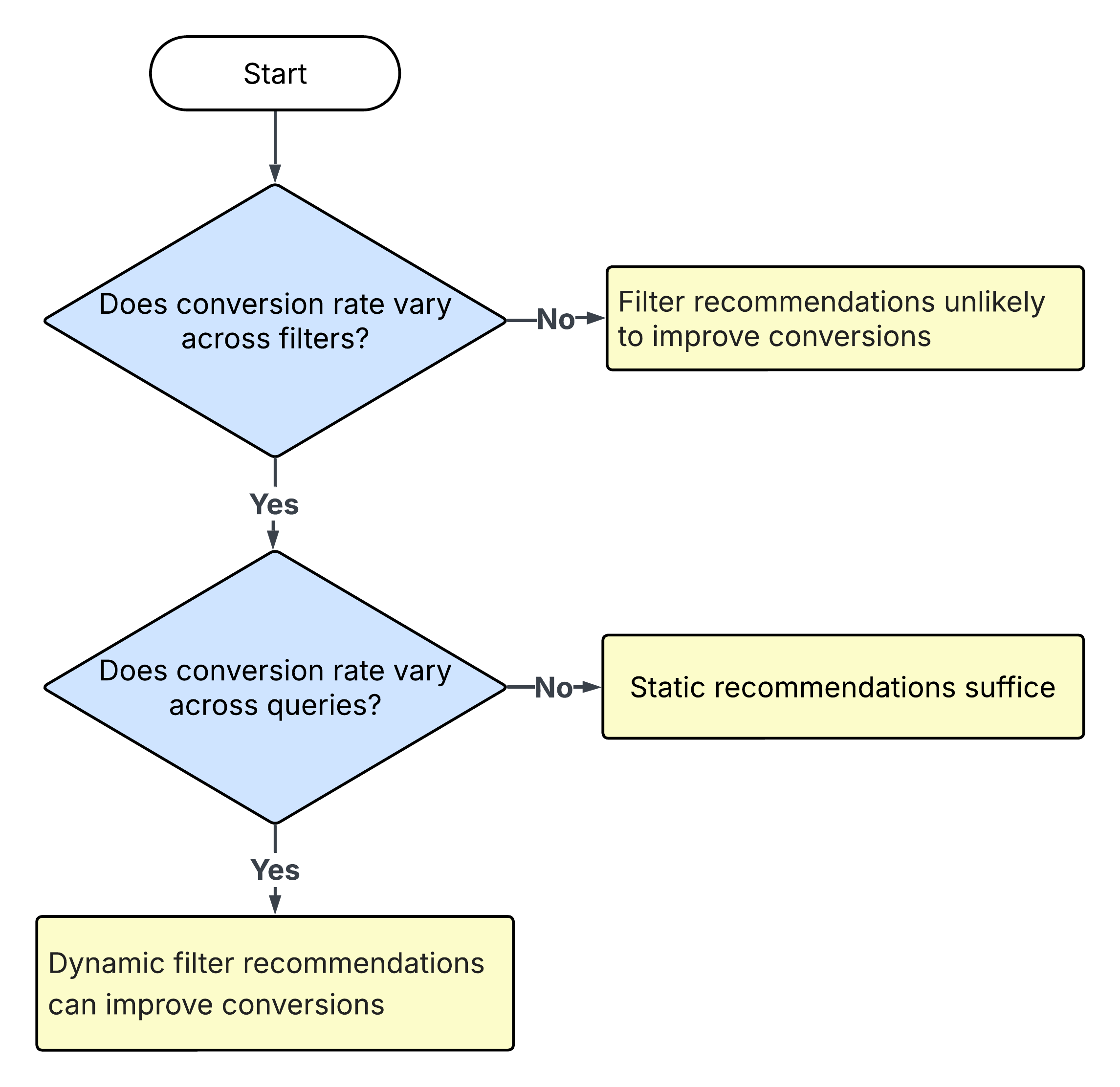}
    \caption{Decision framework for filter recommendation strategies based on conversion rate variability}
    \label{fig:decision-flow}
\end{figure}

To evaluate this hypothesis, we summarize a decision framework, as Figure~\ref{fig:decision-flow} shows, that guides filter recommendation strategy based on Filter Conversion Rate variability. Following this framework, we analyze real-world user behavior data to determine which branch of the decision tree applies to Airbnb's marketplace.

The results, presented in Table~\ref{tab:conversion_rates}, clearly show that Filter Conversion Rates vary across both filters and queries. 
\begin{table}[H]
    \centering
    \resizebox{\columnwidth}{!}{%
    \begin{tabular}{|p{0.35\columnwidth}|p{0.35\columnwidth}|p{0.35\columnwidth}|p{0.35\columnwidth}|}
        \hline
        \textbf{Longer-term stays} & \textbf{Group trips} & \textbf{Short lead time} & \textbf{Long lead time} \\
        \hline
        1.Dedicated workspace & 
        1.Free parking & 
        1.Instant book & 
        1.Free cancellation \\
        2.Washer & 
        2.Hot tub & 
        2.Self check-in & 
        2.Free parking \\
        3.Wifi & 
        3.Barbeque area & 
        3.Free parking & 
        3.Self check-in \\
        \hline
    \end{tabular}%
    }
    \caption{The top filters by Filter Conversion Rates}
    \label{tab:conversion_rates}
\end{table}
\vspace{-7mm}
For longer-term stay queries (number of nights >= 28), the top-performing filters are "dedicated workspace," "washer," and "wifi," which are essential for living and potentially working remotely for an extended period. For group trip queries, the most relevant filters are "free parking," "hot tub," and "barbeque area," offering conveniences and enjoyable group activities. Additionally, the top filters at the two extremes of the lead time spectrum align intuitively with guest needs: the "instant book" filter facilitates immediate bookings for short lead-time queries without requiring host approval, while the "free cancellation" filter provides peace of mind for guests planning further ahead.

The observed variability in Filter Conversion Rates indicates the potential of dynamic filter recommendations to improve booking conversions.

\section{Modeling Framework}
Building on the insights gained in Section 3, we moved forward to formulate the problem, construct training data and train the model, as outlined below. While our focus is on Airbnb, the framework is broadly applicable to other platforms. The core concepts -- searches issued, filters applied, and conversions realized—are fundamental to online marketplaces and e-commerce.

\subsection{Problem Formulation}

Let k be the total number of filter candidates. We denote the filter usage of a search as
\begin{equation}
    F = (F_1, F_2, \dots, F_k), \quad F_i \in \{0, 1\}
    \label{eq:filter-state}
\end{equation}
where $F_i$ =1 indicates that the filter is applied, and $F_i$ =0 indicates it is not.

A special case where no filters are applied is denoted as:
\begin{equation}
    F_\text{none} = (0, 0, \dots, 0)
    \label{eq:filter-state-none}
\end{equation}
Recall that filters’ role in the search journey can be treated as a two step process: (1) guests decide whether to apply certain filters or not and (2) guests choose whether to book a listing from the filtered result set. The first step focuses on filter engagement probability, i.e. given a search query, how likely certain filters will be applied: $P(F | Q)$.

The second step focuses on filter booking probability, i.e. given a search query and certain filters applied, how likely a listing in the filtered results will be booked: $P(B=1 | F, Q)$.

To improve booking conversions, recommended filters must outperform the baseline case (no filter applied) by achieving a higher booking probability:
\begin{equation}
    P(B = 1 \mid F, Q) > P(B = 1 \mid F_\text{none}, Q)
    \label{eq:filter-compare-with-none}
\end{equation}
The booking conversion gain, $g(F)$, measures the improvement in booking probability when certain filters $F$ are applied, compared to the baseline. This improvement is scaled by $P(F \mid Q)$, the likelihood of the filters being applied:
\begin{equation}
    g(F) = P(F \mid Q) \cdot \big[P(B=1 \mid F, Q) - P(B=1 \mid F_\text{none}, Q)\big]
    \label{eq:filter-gain}
\end{equation}
This decomposes the core problem into modeling two probabilities: $P(F | Q)$ and $P(B=1 | F, Q)$

\subsection{Training Data}
Building on the modeling tasks outlined in Section 4.1, we now focus on constructing training data using historical search and booking logs.

\subsubsection{Training Example}
\begin{figure}[t]
    \centering
    \includegraphics[width=1.0\linewidth]{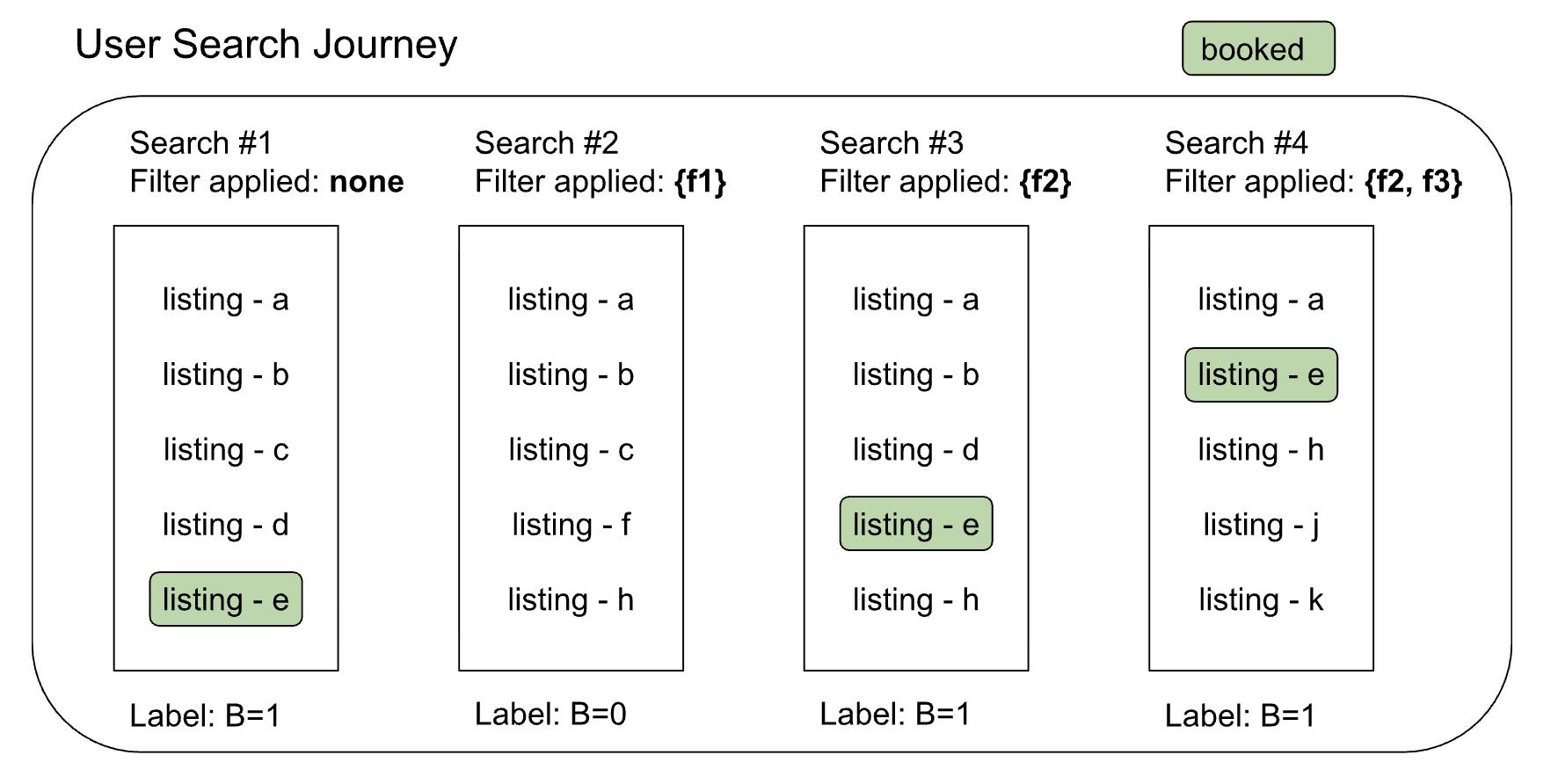}
    \caption{Illustration of filter usage and booking attribution within an example Airbnb user search journey. The search journey is analyzed retrospectively to attribute the booking to preceding searches. Since listing e was booked, Search \#2 is labeled negative (B=0) as the filter applied excluded the booked listing. Searches \#1, \#3 and \#4 are labeled positive (B=1) as their respective filters applied all included the booked listing.}
    \label{fig:label-attribution}
\end{figure}
We construct training examples using one year of de-identified search and booking data, as illustrated in Figure~\ref{fig:label-attribution}. Search logs provide queries, applied filters, and shown listings, while booking logs identify booked listings (listings users ultimately chose to book). We reconstruct the user search journey by joining both logs. By retrospectively analyzing this journey, we examine how filters applied during preceding searches influenced the discoverability of the listings that were booked. Filters can either exclude booked listings from search results (e.g., filter f1 in search \#2 in Figure~\ref{fig:label-attribution}) or elevate booked listings by ranking them higher in the search results for easier discovery (e.g., filters f2 and f3 in searches \#3 and \#4 in Figure~\ref{fig:label-attribution}). 

To improve booking conversions, the model should learn to recommend filters with higher filter booking probabilities than the baseline of no filters applied, as outlined in Equation~\ref{eq:filter-compare-with-none}. To achieve this, the training dataset includes all searches performed by guests, encompassing searches with no filters applied (e.g. search \#1 in Figure~\ref{fig:label-attribution}) and those resulting in no bookings (e.g. search \#2 in Figure~\ref{fig:label-attribution}). This ensures the model learns the relationship between filter usage and booking conversions—whether applying filters boosts or diminishes booking probability compared to the baseline.

We attribute any booking that is not cancelled after a certain period (M days) to any search issued on the same day or within a lookback window of N days before the booking is made. Additionally, the booked listing must be shown in the attributed search’s results. There are around 10 million training examples per day. 

\subsubsection{Training Labels}
For $P(F | Q)$, training labels are defined as whether certain filters are applied for a given search query. If a filter is applied, $F_i$ =1; otherwise, $F_i$ =0. This task is modeled as multi-label classification.
For $P(B=1 | F, Q)$, training labels are defined as whether a booking is attributed to the given search query and filters applied. The label is 1 if there is an attributed booking, otherwise the label is 0. The attribution process is presented in Section 4.2.1 and illustrated in Figure~\ref{fig:label-attribution}.

\begin{table}[t]
    \centering
    \resizebox{\columnwidth}{!}{%
    \begin{tabular}{p{0.5\columnwidth}|p{0.30\columnwidth}|p{0.20\columnwidth}}
        \toprule
        \textbf{Feature} & \textbf{Feature category} & \textbf{Data type} \ \\
        \midrule
        Location searched (unique ID) & Query feature & Categorical \\
        Number of adults & Query feature & Continuous \\
        Number of children & Query feature & Continuous \\
        Number of infants & Query feature & Continuous \\
        Check-in: month of the year & Query feature & Cyclical \\
        Check-in: day of the month & Query feature & Cyclical \\
        Check-in: day of the week & Query feature & Cyclical \\
        Check-out: month of the year & Query feature & Cyclical \\
        Check-out: day of the month & Query feature & Cyclical \\
        Check-out: day of the week & Query feature & Cyclical \\
        Number of nights & Query feature & Continuous \\
        Lead time (days before check-in) & Query feature & Continuous \\
        Platform & Query feature & Categorical \\
        Device type & Query feature & Categorical \\
        Filter applied (unique ID) & Filter feature & Categorical \\
        \bottomrule
    \end{tabular}%
    }
    \caption{Model features}
    \label{tab:model_features}
\end{table}

\subsubsection{Model Features}
In introducing a new model rather than refining an existing one, we start with a minimal set of directly observed and reported features \cite{zinkevich2017rules}, prioritizing interpretability and minimizing data dependency debt \cite{sculley2015hidden}. The model features, listed in Table~\ref{tab:model_features}, are derived from search queries explicitly provided by guests. Personalization features have been intentionally deferred to future work, as discussed in Section 7. A feature ablation study is detailed in Section 6.3.3 to evaluate each feature's contribution to the model's performance.

We encode features in Table~\ref{tab:model_features} as follows:
\begin{itemize}
\item {\textbf{Categorical features}}: features are mapped to dense embedding vectors, learned jointly with the model during training.
\item {\textbf{Continuous features}}: features are normalized.
\item {\textbf{Cyclical features}}: For a cyclical feature $x$ with a known period $P$ (e.g., $P = 12$ for month of the year), the feature can be encoded as:
\begin{equation}
\begin{split}
     x_{\text{sin}} = \sin\left(2\pi \frac{x}{P}\right)  \\
     x_{\text{cos}} = \cos\left(2\pi \frac{x}{P}\right) \\
    \label{eq:cyclical-encoding}
\end{split}
\end{equation}
\end{itemize}

\subsection{Model Architecture}
\begin{figure}[t]
    \centering
    \includegraphics[width=0.8\linewidth]{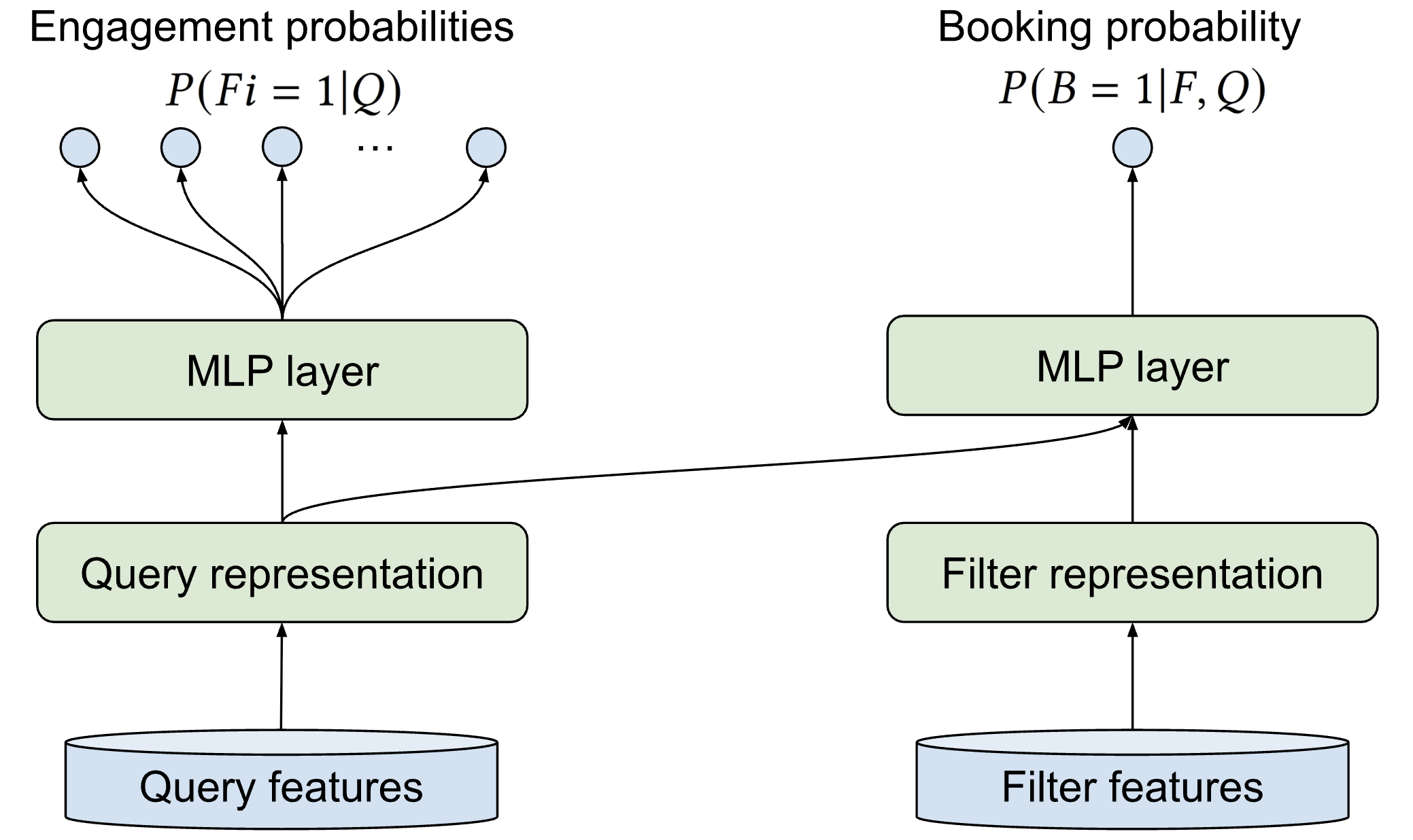}
    \caption{Model architecture}
    \label{fig:model-architecture}
\end{figure}

The model architecture is presented in Figure~\ref{fig:model-architecture}. It incorporates a multi-task learning approach, where two tasks are jointly trained: one is to predict $P(Fi=1 | Q)$ and another task is to predict $P(B=1 | F, Q)$. Both tasks share query and user representation to ensure efficient representation learning. The $P(B=1 | F, Q)$ task takes filter features as an additional input. 

The neural network architecture employs a multi-layer perceptron (MLP) with two layers and hidden sizes of [64, 32]. The loss function is a weighted linear combination of losses from both tasks:
\begin{equation}
    \mathcal{L}_{\text{total}} = \mathbf{w}_{\text{engagement}} \cdot \mathcal{L}_{\text{engagement}} + \mathbf{w}_{\text{booking}} \cdot \mathcal{L}_{\text{booking}}
    \label{eq:loss-function}
\end{equation}

\section{Productionization and Practical Considerations}
This section details the system design and engineering solutions that enabled us to operationalize the framework presented in Section 4 into a production system at Airbnb.

\subsection{Serving Architecture}
\subsubsection{Latency and QPS optimization}
Filter recommendations are generated per query and included in the search response. This requires strict inference latency (<10ms) and high throughput (>1000 QPS). To meet these requirements, we employ two key designs:
\begin{itemize}
\item {\textbf{Model serving}}: Figure~\ref{fig:serving-arch} (right) illustrates model serving via TensorFlow Serving (TFS). We use a sidecar architecture pattern \cite{burns2016design}, co-locating the TFS sidecar with the search service in the same pod. This design eliminates network hops, reducing latency and improving throughput scalability.
\item {\textbf{Feature serving}}: Figure~\ref{fig:serving-arch} (left) shows feature serving. Features serving is shared between the listing search and filter recommendation modules, which has the following benefits:
\begin{enumerate}
    \item Single feature fetch reduces latency.
    \item Feature sharing across modules reduces computation.
    \item Centralized management ensures feature consistency across modules.
\end{enumerate}
\end{itemize}

\subsubsection{Enabling item-aware filter recommendation}
\begin{figure}[t]
    \centering
    \includegraphics[width=1\linewidth]{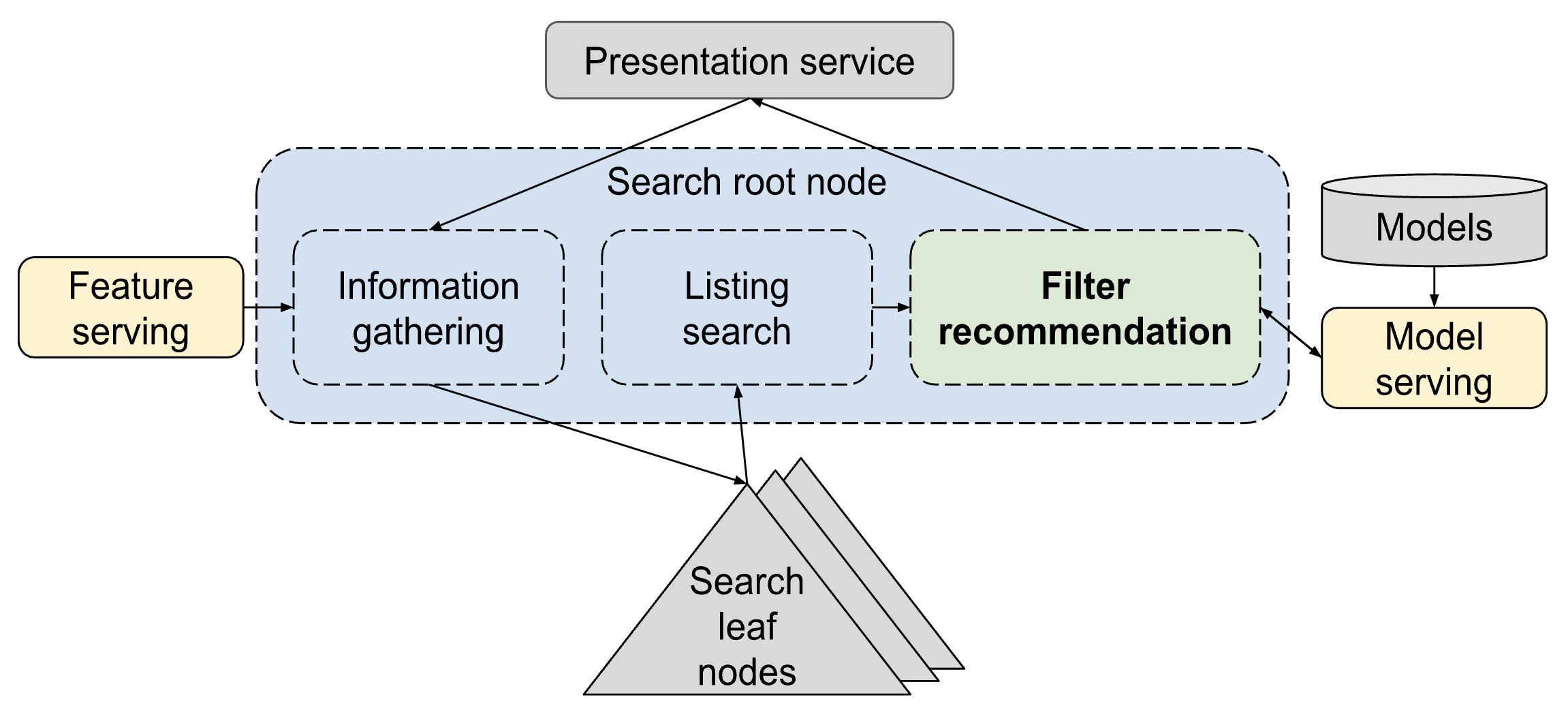}
    \caption{Serving architecture}
    \label{fig:serving-arch}
\end{figure}

Figure~\ref{fig:serving-arch} (middle) shows the \textbf{filter recommendation} module positioned after the \textbf{listing search} module in the Search Root Node. This sequencing allows the filter recommendation module to adjust results using real-time item-level data from the listing search module. For example, we implement a post processing step — remove filter recommendations that will lead to less than 18 listings. Online A/B test results in Section 6.2 show how this post processing reduces searches in low inventory states. Moreover, this setup enables advanced modeling. Item-level data can be seamlessly incorporated into the model, extending $P(B=1 | Q, F)$ to $P(B=1 | Q, F, I)$, where $I$ represents items. We explore this possibility as future work in Section 7.

\subsection{Cold Start}
Building a new ML-based recommendation system often faces cold-start challenges such as a lack of historical data and presentation bias. Our approach offers inherent advantages (Section 5.2.1), complemented by an additional empirical trick (Section 5.2.2).

\subsubsection{Leveraging Historical Data Beyond Engagement}
Our approach incorporates booking data to model filter conversion probability $P(B=1|F,Q)$. This allows us to utilize richer historical data, linking filter usage directly to lower-funnel conversions rather than stopping at engagement logs. For instance, the way we generate training examples (Section 4.2) connects filter usage to bookings, leveraging conversion data that most e-commerce or online marketplace platforms already log but often overlook as a basis for recommending filters.

\subsubsection{Mitigating Presentation Bias}
Traditional approaches face significant presentation bias as filters are often ranked using heuristics or fixed orders before introducing a model. By emphasizing $P(F | Q)$, these systems risk skewed results influenced by display prominence. For example, filters ranked higher by heuristics tend to show inflated $P(F | Q)$, while filters ranked lower may appear artificially suppressed. This bias occurs because training data reflects filter usage shaped by non-randomized rankings.

Our approach additionally models $P(B=1|F,Q)$, which is conditioned on filter application and less influenced by presentation placement. This shift ensures rankings are more determined by filter utility (e.g., the relevance and quality of results after applying the filter) than display bias.

To further mitigate bias, we adjusted Equation~\ref{eq:filter-gain} to emphasize the term $P(B=1 | F, Q)$, as it is less influenced by display order. The modified formula introduces N as a parameter to control the weight of the filter booking probability term:
\begin{equation}
    rankingScore = P(F | Q)*P(B=1 | F, Q)^N
    \label{eq:filter-ranking-score}
\end{equation}
Increasing N amplifies the influence of the booking term in the ranking process. Section 6.3 demonstrates the effectiveness of this adjustment through online A/B tests.

\section{Experiments and Results}
We evaluate the models offline to assess their predictive power in Section 6.1, conduct online A/B tests to validate our key hypothesis -- recommending search filters can drive conversions in Section 6.2., and perform ablation study to validate the effectiveness of key designs of our approach in Section 6.3.

\begin{table}[t]
    \centering
    \resizebox{\columnwidth}{!}{%
    \begin{tabular}{p{0.3\columnwidth}|p{0.3\columnwidth}|p{0.3\columnwidth}}
    \toprule
        \textbf{Approach} & \textbf{PR-AUC} & \textbf{Improvement} \\
    \midrule
        Production & 0.04795 & baseline \\
        Statistics-based & 0.06821 & +42.2\% \\
        ML-based & 0.09613 & +100.5\% \\
    \bottomrule
    \end{tabular}%
    }
    \caption{Offline evaluation of booking-conversion prediction based on filter usage}
    \label{tab:offline_evaluation_results}
\end{table}

\subsection{Offline Evaluation}
Given that the problem is decomposed into two binary classification subproblems with highly imbalanced datasets (with the positive class being significantly outnumbered by the negative class),  we use PR-AUC (Area Under the Precision-Recall Curve) as the offline evaluation metric to evaluate the quality of our binary classifiers. 

Below, we focus on evaluating the $P(B=1 | F, Q)$ model, newly introduced in this paper to improve conversion through filter recommendations. The evaluation dataset consists of search queries, filter usage, and booking outcomes as labels.

We compare three approaches, detailed below, with results summarized in Table~\ref{tab:offline_evaluation_results}:

\begin{itemize}
\item {\textbf{Production (baseline)}}: The production system previously deployed at Airbnb serves as the baseline. It computes the "necessity" of a filter as a prediction -- specifically, whether the booked listing satisfies the filter if the filter was ever applied during the search journey. Unlike our proposed approach, this baseline does not require the booking to result from a filtered search, thus capturing a correlational rather than causal link.

\item {\textbf{Statistics-based}}: This approach computes Filter Conversion Rates as predictions (see Equation~\ref{eq:filter-conversion-rate}). FCR is generated using a 7-day window for a given filter and destination combination. This follows the modeling framework presented in Section 4 but relies solely on historical statistics rather than ML predictions.

\item {\textbf{ML-based}}: This approach trains a ML model to predict $P(B=1 | F, Q)$, as detailed in Section 4. The model captures complex patterns in the data, offering a more robust and adaptive solution compared to Statistics-based.
\end{itemize}

The results provide several key insights:
\begin{enumerate}
    \item Feasibility: Search queries and their filter usage are predictive of booking conversions, as evidenced by improvements in PR-AUC with more advanced approaches.
    \item Necessity of ML models: The underlying patterns are sufficiently complex to justify the development of dedicated ML models, supported by the performance gap between the Statistics approach and the ML-based approach.
    \item Potential for conversions: There is significant potential for improving booking conversions, demonstrated by the substantial performance gap between the Production approach and the new ML-based approach.
\end{enumerate}

\begin{table}[t]
    \centering
    \resizebox{\columnwidth}{!}{%
    \begin{tabular}{p{0.2\columnwidth}|p{0.2\columnwidth}|p{0.2\columnwidth}|p{0.2\columnwidth}|p{0.2\columnwidth}}
    \toprule
        \textbf{Application} & \textbf{Uncanceled bookings} & \textbf{Booked listing in filtered results} & \textbf{Searches using filters} & \textbf{Searches low inventory}\\
    \midrule
        filter panel & +0.28\%* & +10.2\%*** & +9.8\%*** & -0.81\%*** \\
        filter bar & +2.4\%* & +8.1\%** & +6.4\%* & -1.4\% \\
    \bottomrule
    \end{tabular}%
    }
    \caption{Online A/B test results showing the relative lift of our approach over the production baseline across two surfaces (significance: * p < 0.05; ** p < 0.01; *** p < 0.001)}
    \label{tab:online_results}
\end{table}

\subsection{Online A/B Tests}
To validate our approach in a real-world setting, we conducted online A/B tests on Airbnb’s search platform.

\textbf{Application 1: Filter Panel}. The first application is recommending amenity filters within the filter panel (see Figure~\ref{fig:recommended-amenities} for the UI). In this panel, the amenity filter section features a subsection titled “Popular”, where six amenity filters are recommended.

The A/B test compared two groups: the control group received filter recommendations generated using the Production approach, while the treatment group received filter recommendations generated by the ML-based approach (as outlined in Section 6.1 and Table~\ref{tab:offline_evaluation_results}). The results, summarized in the first row of Table~\ref{tab:online_results}, validated the effectiveness of our approach:
\begin{enumerate}
    \item Uncanceled bookings increased by 0.28\%: This validated our key hypothesis that booking conversions can be improved by recommending search filters.
    \item Searches using filters increased by 9.8\% and booked listings in filtered results increased by 10.2\%: These metrics show improvements throughout the funnel. First, recommendations successfully drove higher filter adoption; second, filtered results led to more bookings.
    \item Searches hitting low inventory state decreased by 0.81\%: This improvement stems from the post-processing step in the serving pipeline (Section 5.1.2), where the recommendation module uses facet counts from the listing search module to exclude filters leading to low-inventory states.
\end{enumerate}

\textbf{Application 2: Filter Bar}. The second application is recommending filters within the filter bar (see Figure~\ref{fig:filter-bar} for the UI). The filter bar allows users to apply filters more conveniently for selected markets, eliminating the need for extra clicks and scrolls to access the filter panel.

The same approach is applied, where $P(F | Q)$ and $P(B=1 | F, Q)$ are modeled and combined to rank filters. The results, shown in the second row of Table~\ref{tab:online_results}, also show an increase in uncanceled bookings, with a lift of + 2.4\%. The larger impact compared to the first application can be attributed to the more prominent placement, the broader variety of filters available for recommendation, and the ease of applying recommended filters. Recommended filters are displayed directly in the dedicated filter bar, encompassing more than just amenity filters, and requiring only a single click to apply. The significant increases in searches using filters and booked listings in filtered results further reinforce that the booking conversion lift is driven by improved filter recommendations.

\subsection{Ablation Study}
Thus far, we have demonstrated the effectiveness of our proposed approach. To ensure rigor and highlight the critical contributions of our design choices, we conducted an ablation study to evaluate their individual impacts on the overall success.

\textbf{Effectiveness of Problem Formulation}.
The problem formulation in Section 4.1 introduces the filter conversion probability term $P(B=1|F,Q)$, differentiating our approach from prior work that focuses solely on filter engagement $P(F|Q)$. To validate this contribution, we ran online A/B tests comparing the baseline (ranking filters purely by $P(F|Q)$) with the treatment (ranking filters by a combination of $P(F|Q)$ and $P(B=1|F,Q)$). These tests were reproduced across two distinct markets, with results presented in Table~\ref{tab:online_results_ablation}.

\begin{table}[t]
    \centering
    \resizebox{\columnwidth}{!}{%
    \begin{tabular}{p{0.2\columnwidth}|p{0.4\columnwidth}|p{0.4\columnwidth}}
        \toprule
        \textbf{Market} & \textbf{Ranking formula} & \textbf{Uncanceled bookings}\\
        \midrule
        X & $P(F|Q)$ only & -0.59\%  \\
         & $P(F|Q)$ and $P(B=1|F,Q)$ & +2.9\%* \\
        \midrule
        Y & $P(F|Q)$ only & -0.08\% \\
         & $P(F|Q)$ and $P(B=1|F,Q)$ & +0.76\%* \\
        \bottomrule
    \end{tabular}%
    }
    \caption{Online A/B tests results for ablating $P(B=1|F,Q)$ (* indicates p-value < 0.05)}
    \label{tab:online_results_ablation}
\end{table}

\begin{table}[t]
    \centering
    \resizebox{\columnwidth}{!}{%
    \begin{tabular}{p{0.2\columnwidth}|p{0.4\columnwidth}|p{0.4\columnwidth}}
        \toprule
        \textbf{value of N} &\textbf{Ranking formula} & \textbf{Uncanceled bookings}\\
        \midrule
        N=2 & $P(F | Q)*P(B=1 | F, Q)^2$ & +0.28\%* \\
        N=0.5 & $P(F | Q)*P(B=1 | F, Q)^{0.5}$ & +0.19\% \\
        \bottomrule
    \end{tabular}%
    }
    \caption{Online A/B tests results for cold start strategy (* indicates p-value < 0.05)}
    \label{tab:online_results_cold_start}
\end{table}

The results confirm that the conversion lifts (uncanceled bookings) are driven by our new problem formulation (Section 4.1), specifically the incorporation of $P(B=1|F,Q)$.

An additional question arises: if $P(B=1|F,Q)$ drives conversions so effectively, is $P(F|Q)$ still necessary? The answer is yes. Relying solely on $P(B=1|F,Q)$ to rank filters can result in a suboptimal user experience, as the recommended filters may seem irrelevant to the query. This is because $P(B=1|F,Q)$ is conditioned on the already applied filters $F$. We implemented ranking filters solely based on $P(B=1|F,Q)$ but found that it failed to meet the user experience standards required for advancing to online A/B testing. This was because the model neglected the query-relevance signal from filter engagement $P(F|Q)$, resulting in some seemingly irrelevant filter recommendations.

\textbf{Effectiveness of Cold Start Strategy}.
Section 5.2 addresses presentation bias through the use of a larger $N$ to place more weight on filter conversion, as defined in Equation \ref{eq:filter-ranking-score}. To validate this adjustment, we conducted online A/B tests with two configurations: $N=0.5$ (smaller weight) and $N=2$ (larger weight), keeping all other settings constant. As presented in Table~\ref{tab:online_results_cold_start}, the variant with $N=2$ demonstrated a statistically significant lift in uncanceled bookings, while the $N=0.5$ variant exhibited regression in the same metric and became statistically insignificant. These results highlight the importance of putting a larger weight on filter conversion probabilities to mitigate presentation bias and drive conversion gains.

\textbf{Feature Ablation Analysis}.
To build the initial version of the ML model, we curated a minimal yet impactful set of features directly observed to contribute to a robust model foundation for future iterations \cite{zinkevich2017rules}. The selected features are listed in Table~\ref{tab:model_features} in Section 4.2. To assess the contribution of each feature, we conducted a feature ablation study, with results summarized in Table~\ref{tab:feature-ablation}. Each model variant is trained after entirely removing a specific subset of features. The significant drop in performance across both tasks observed during ablations confirms that the feature set is well-suited as a starting point.
\begin{table}[t]
    \centering
    \resizebox{\columnwidth}{!}{%
    \begin{tabular}{l|cc|cc}
        \toprule
        \textbf{Model / Feature Set Removed} & \textbf{$P(F|Q)$} & \textbf{$\Delta$ (\%)} & \textbf{$P(B=1|F,Q)$} & \textbf{$\Delta$ (\%)} \\
        \midrule
        \textbf{Full Model (Baseline)} & \textbf{0.4836} & - & \textbf{0.0961} & - \\
        \midrule
        - Location Searched & 0.3753 & -22.4 & 0.0759 & -21.0 \\
        - Dates & 0.4570 & -5.5 & 0.0764 & -20.5 \\
        - Guest Counts & 0.4573 & -5.4 & 0.0893 & -7.1 \\
        - Platform \& Device & 0.4712 & -2.6 & 0.0853 & -11.2 \\
        \bottomrule
    \end{tabular}%
    }
    \caption{Feature ablation on PR-AUC metrics. ``Location Searched'' is the most important feature set, followed by ``Dates'', with the remaining sets having a smaller but still significant impact.}
    \label{tab:feature-ablation}
\end{table}

\section{Conclusions and Extensions}
We presented how conversion-oriented filter recommendation was built from zero to one at Airbnb. We first introduced a modeling \textbf{framework} aimed at improving conversions by recommending search filters.  The proposed framework was successfully operationalized into a production \textbf{system}, tackling practical problems like cold start and efficient serving architecture. Finally, offline and online evaluations \textbf{validated} the system's ability to drive conversions. 
\subsection{Model Extensions}
Our framework can be extended in several key directions:
\begin{itemize}
    \item Item-Aware Recommendations: Extend the model from $P(B=1 | Q, F)$ to $P(B=1 | Q, F, I)$, where I represents item-level features. The serving architecture (Section 5.1) positions the filter recommendation module after the listing search module, enabling access to real-time item-level data for enhanced recommendations.
    \item Leveraging Presentation Data: Our current approach cold-starts filter ranking models using historical data, without incorporating presentation data, such as whether a filter was displayed as recommended. With sufficient presentation data in the future, we can explore conditioning $P(F | Q)$ and $P(B=1 | F, Q)$ on the impressions of recommended filters. This would refine the model beyond historical data and better adapt it to user behavior in the new UIs.
    \item Beyond Qualitative Filters: While the current modeling framework supports qualitative filters (e.g. free parking:true), extending the approach to numeric filters (e.g. max price: 100) remains an open question for future research.
\end{itemize}

\begin{figure}[t]
    \centering
    \includegraphics[width=0.37\linewidth]{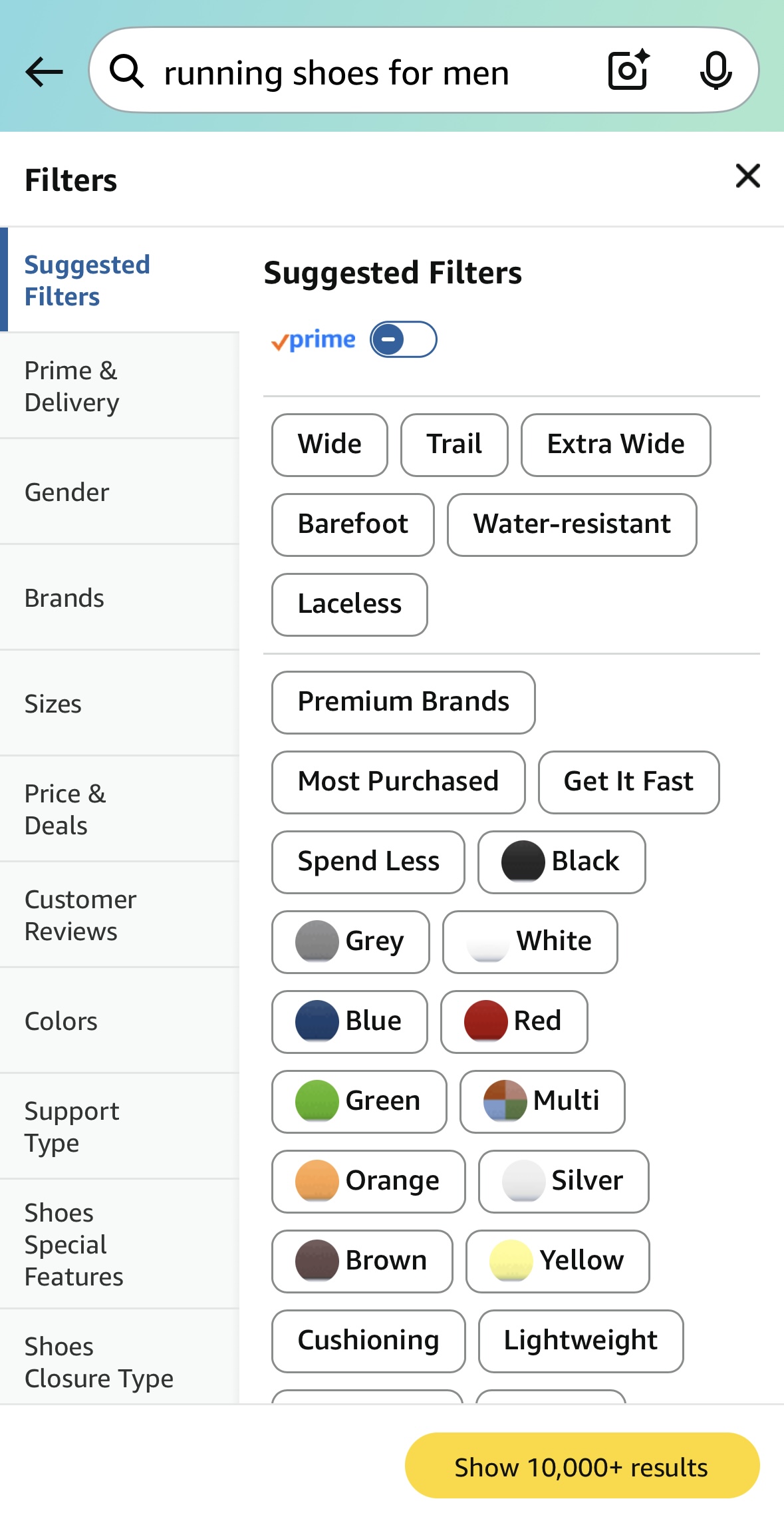}
    \caption{Filter interface on Amazon, providing a dynamic ``Suggested Filters'' section}
    \label{fig:filter-amazon}
\end{figure}

\begin{figure}[t]
    \centering
    \includegraphics[width=0.37\linewidth]{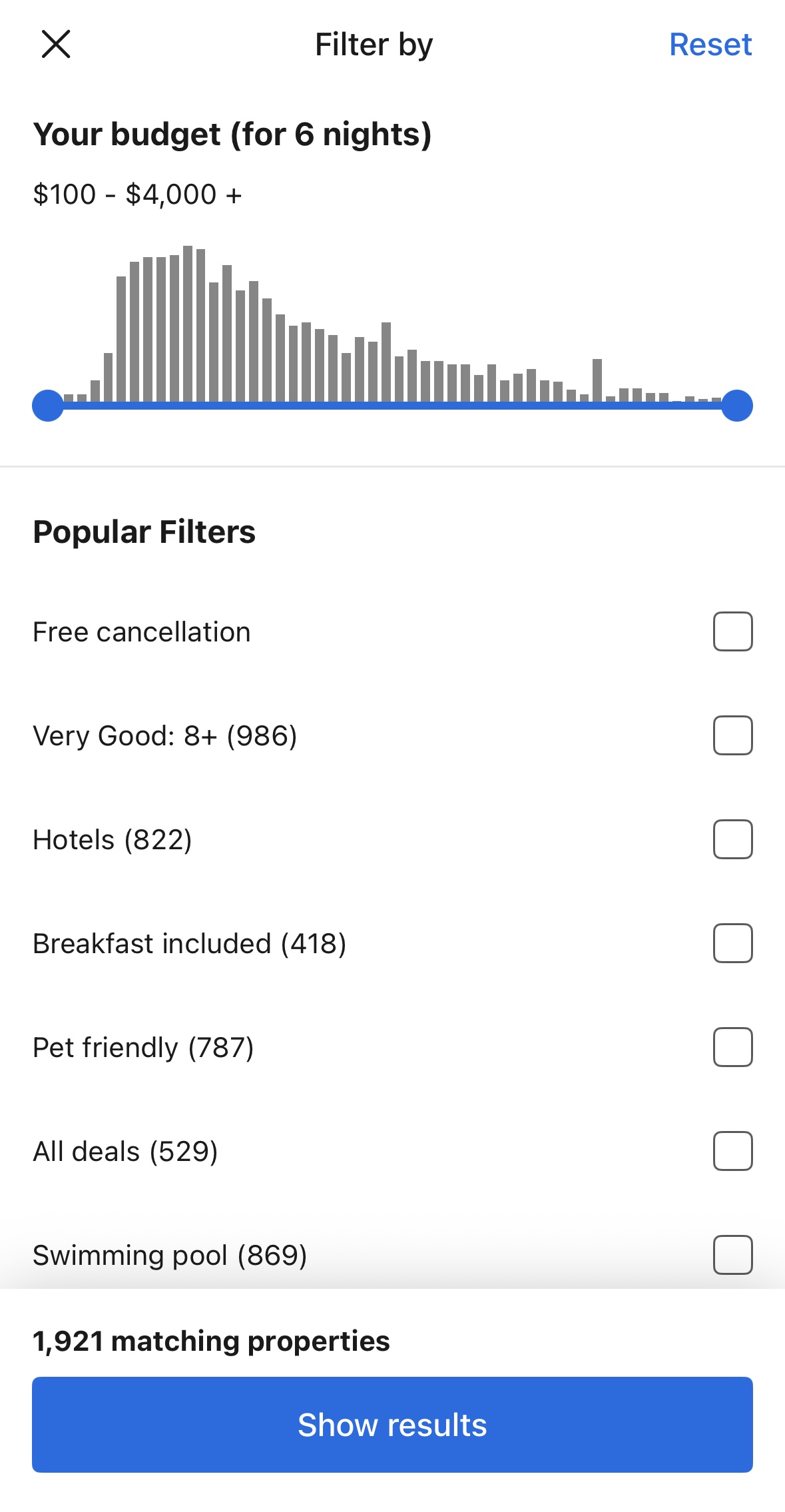}
    \caption{Filter interface on Booking.com, providing a dynamic ``Popular Filters'' section}
    \label{fig:filter-booking}
\end{figure}

\begin{figure}[t]
    \centering
    \includegraphics[width=0.37\linewidth]{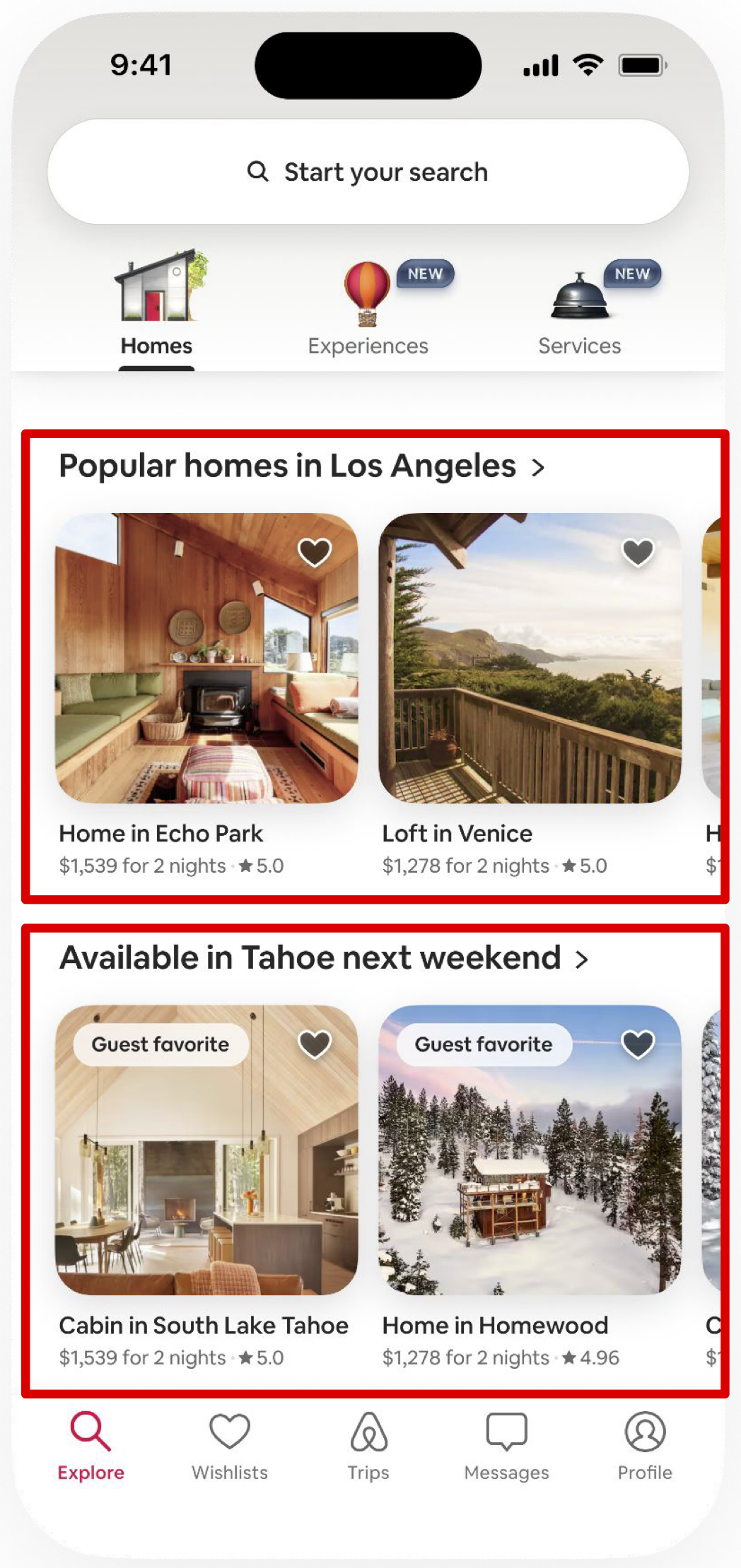}
    \caption{Carousels on Airbnb homepage}
    \label{fig:homepage-carousel}
\end{figure}

\subsection{Broader Applicability}
While this paper focuses on Airbnb, the framework is domain-agnostic and can be extended to other platforms and search tools. The generalizability of our framework stems from the decomposition of the problem into two widely applicable modeling tasks:
\begin{itemize}
    \item Modeling Tool Engagement: Predicting the probability of user interaction with a search tool based on their query ($P(F | Q)$). This captures immediate user interest and perceived relevance.
    \item Modeling Tool Utility: Predicting the probability of a downstream conversion given the use of that tool ($P(B=1 | Q, F)$). This captures the tool's utility in achieving the platform's business objective.
\end{itemize}

\textbf{Generalizing to Other Platforms}.
The framework is applicable to other online marketplaces that use filters to guide conversions, including e-commerce platforms like Amazon (Figure~\ref{fig:filter-amazon}), travel sites like Booking.com (Figure~\ref{fig:filter-booking}) and food delivery app like DoorDash.

\textbf{Generalizing to Other Search Tools}.
The framework can also be applied to other dynamic search and discovery tools by modeling and balancing engagement against utility.
\begin{itemize}
    \item Carousels: The framework can rank themed carousels (Figure~\ref{fig:homepage-carousel}) to prioritize those that effectively lead to conversions, not just attract clicks or swipes on carousels.
    \item Choice Chips: Modern interfaces often use "choice chips" or visual toggles instead of traditional filter lists (e.g. buttons for "Restaurant," "Cafe," "Hotels" on a map). The framework can inform the dynamic ordering of these chips, balancing their engagement potential with their utility in guiding users to conversions.
\end{itemize}

\begin{acks}
We owe the success of this project to the incredible cross-functional collaboration between teams. Our sincere thanks go to the Search Input team (Alejandro Erviti, Shuai Chen, Stella Zhang, Amy Odena, Sergii Rudenko, Julian Adams, Carol Leung, Rachel Zhao), Product team (Clarence Quah, Judith Dito), Design team (Elva Chen, Adam Shutsa), Search Backend team (Karen Ouyang, Soumyadip Banerjee, Yangbo Zhu), Global Markets Data Science team (Momo Hou) and Advanced Analytics team (Matt Yun).

\end{acks}

\bibliographystyle{ACM-Reference-Format}
\bibliography{references}

\end{document}